\documentstyle[psfig]{mn}

\title{Broad emission lines from opaque electron-scattering 
 environment of SN~1998S
 }
 
\author[N.N. Chugai]
  { N.N. Chugai \\
 Institute of Astronomy, RAS, Pyatnitskaya 48, 109017 Moscow, Russia}

\date{Accepted 2001
      Received 2000;
      in original form 2000}

\pagerange{\pageref{firstpage}--\pageref{lastpage}}
\pubyear{2001}

\begin{document}

\maketitle

\label{firstpage}

\begin{abstract}

I propose that broad narrow-topped emission lines with full 
 width at zero intensity  $>20000$ km s$^{-1}$ revealed by 
 early-time spectra of SN~1998S originate from a dense 
 circumstellar (CS) shell with the outer 
 radius $R\approx 10^{15}$ cm.
The tremendous line width is the result of 
 multiple scattering of photons of the narrow line  
 by thermal electrons of the opaque CS gas. 
The H$\alpha$ line on 1998 March 6 is reproduced 
 by Monte Carlo simulations provided 
 the Thomson optical depth of the CS shell is  
 3--4 at this epoch.
The high density of CS shell implies that a cool dense shell (CDS), 
 which forms at the ejecta-wind interface, is opaque in the Paschen continuum
 for about 40--50 days after the explosion.
This explains, why strong lines from the supernova ejecta were not seen 
 during an initial period of about 40 days after the discovery.
The Thomson optical depth recovered from the electron-scattering wings 
 is consistent with observational constraints imposed by 
 the bolometric luminosity, photospheric radius and expansion velocity 
 of the CDS.

\end{abstract}


\section{Introduction}

Supernova SN~1998S discovered on March 2.68 UT 
 (Li \& Wan 1998) belongs to a diverse family of SN~IIn, 
 which are generally believed to interact with a dense 
 presupernova wind (Filippenko 1991, 1997; Chugai 1990).
Of particular interest are the earliest spectra of
 SN~1998S, taken before 1998 March 12, with 
 strong Balmer emission lines and Wolf-Rayet (WR) features, 
  N~III 4640 \AA, He~II 4686 \AA\, on a smooth black-body continuum
 (Leonard et al. 2000; Liu et al. 2000; Fassia et al. 2001).
The lines have similar profiles, composed of an unresolved line with 
 full width at half maximum (FWHM) $<300$ km s$^{-1}$ 
 and a broad line with full width near zero intensity 
  $\geq 20000$ km s$^{-1}$ (Leonard et al. 2000).
The narrow component is identified with the CS gas,
 while the broad line is attributed to the SN ejecta 
 (Leonard et al. 2000), in accord with a picture generic to SN~IIn.

I will present quite different view on the origin of  
 the broad component.
It will be explained as 
 a result of broadening of the narrow line due to multiple scattering 
 by thermal electrons in the opaque CS gas.
I will show that this conjecture is
 consistent with  observational constraints imposed 
 by the bolometric light curve, photospheric radius and 
 velocity of the ejecta-wind interface.
I first outline the phenomenological picture, 
  which will hint, why the CS gas and not SN ejecta is
  an appropriate site for the early-time broad emission (section 2),
  then calculate simple models of the ejecta deceleration 
 and bolometric light curve (section 3).
In section 4 the line profile will be simulated using Monte Carlo 
 technique.

Henceforth the distance 17 Mpc, the extinction $A(V)=0.68$ 
 (Fassia et al. 2000), and  the redshift 850 km s$^{-1}$ 
 (Fassia et al. 2001) are assumed.

\section{The phenomenological model}

Bellow we outline the qualitative physical picture 
 of the spectrum formation in the early-time SN~1998S using 
 arguments based upon the available observational data and 
 results obtained in other papers.

\subsection{Why CS Thomson scattering?}

SN~1998S exploded after 1998 February 23.7 (Leonard et al. 2000),
 which suggests that on March 6 SN was less then ten days old.
At this age SN ejecta is essentially opaque, and  
 the emission lines from SN ejecta 
 must be affected by the occultation, {\em i.e.} they must show 
  strong line asymmetry with the suppressed red part.
The early emission lines, on the contrary,  are symmetric  
 (Leonard et al. 2000), and their origin from SN~1998S ejecta 
 is, therefore, unlikely.
 
The alternative explanation for the broad lines, we propose,
 is that they are the result of narrow line broadening
  due to electron scattering in an opaque CS shell outside the 
  SN photosphere.
The following simple arguments give an order of magnitude estimate of the 
 required Thomson optical depth of the CS shell to produce the 
 broadening effect. 
Let the H$\alpha$ emissivity and scattering electrons are 
 homogeneously distributed in a spherical layer with the Thomson
 optical depth $\tau_{\rm T}$.
The expected ratio of the unscattered, {\em i.e.} narrow component, to 
 the total emergent line flux is then
 $U\approx [1- \exp(-\tau_{\rm T})]/\tau_{\rm T}$. 
On the other hand, assuming that the CS shell velocity is about 
 $10^3$ km s$^{-1}$,
 which is hinted by the UV lines in HST spectra on March 16 
 (Lentz et al. 2001), we obtain that the narrow component 
 (or unscattered emission) of H$\alpha$ in the spectrum
  on March 6 (Fassia et al. 2001) comprises roughly $20-25\%$ of the 
 overall line emission, which suggests $U(\tau_{\rm T})\approx 0.2-0.25$.
With this value the above expression for $U(\tau_{\rm T})$ implies    
 the optical depth of the CS gas $\tau_{\rm T}\approx 4$. 
More precise value of $\tau_{\rm T}$ will be obtained in Section 4.  

\subsection{The distribution of the CS matter} 
 
The abrupt disapearence of the early-time emission lines after 
 March 12 means that a dense CS shell, responsible for these lines,  
 was overtaken by the SN ejecta to this moment. 
This argument was used by Fassia et al. (2001) to estimate the 
 outer radius of the CS shell, $R_{\rm c,1}\sim 10^{15}$ cm.
Note, the radius of the photosphere on March 6 estimated by the 
 extrapolation of empirical data provided by Fassia et al. (2000) is 
 $R_{\rm p}\approx 5\times 10^{14}$ cm $<R_{\rm c,1}$.
 
Apart from this inner CS shell, there is also the CS matter in the 
 outer region $r> R_{\rm c,1}$,
 which is revealed by narrow absorption and emission lines in spectra 
 seen after March 12 and by the emergence 
 at $t> 50$ d of the 
 'boxy' components of H$\alpha$ and He I 10830 \AA\ caused by the 
 interaction (Fassia et al. 2001).
The modeling of UV lines in the HST spectrum on March 16 (Lentz et al. 2001)
 indicates that the Thomson optical depth of the CS gas in the region
  $r>10^{15}$ cm is only $\tau_{\rm T}\approx 0.2$.
Therefore, parctically all the optical depth $\tau_{\rm T}\approx 4$ should 
 be accumulated in the range of $R_{\rm p}<r<R_{\rm c,1}$ with 
 a linear density $w= 4\pi r^2 \rho =
 4\pi R_{\rm p}(\tau_{\rm T}/k_{\rm T})/(1-R_{\rm p}/R_{\rm c,1})
  \approx 2\times 10^{17}$ g cm$^{-1}$, where we assume $\rho \propto r^{-2}$, 
  and $k_{\rm T}=0.3$ cm$^2$ g$^{-1}$.
The estimated value of $w$ is tremendous, but yet is in line with 
 the similar records in other SN~IIn (Chugai 1992).
As to the outer CS gas, the value $\tau_{\rm T}\approx 0.2$ in the 
 range $r>R_{\rm c,1}$ suggests that 
 $w=4\pi R_{\rm c,1}\tau_{\rm T}/k_{\rm T}\sim 10^{16}$ g cm$^{-1}$, 
 which is consistent with the wind density required by  
 the late-time ($t> 1$ yr) radio and X-ray emission (Pooley et al. 2001). 

The above analysis implies that the CS environment has 
 a 'dense core plus rarefied halo' morphology. 
This structure  will by described here  
 as a superposition of two components ($j=1,\, 2$), each described 
 by the expression
 
\begin{equation}
\rho_j=\rho_{{\rm c},j}(r/R_{\rm c,j})^{s_j}[1+(r/R_{\rm c,j})^{p_j}]^{-1}.
\label{eq:cs}
\end{equation}   

\noindent For the template model of the CS gas distribution 
 (Fig. 1), which meets above requirements 
 and will be used bellow for the light 
 curve and line profile models,
 we adopt $s_1=-2$, $p_1=10$, $R_{\rm c,1}=8\times 10^{14}$ cm,
 $s_2=0$, $p_2=2$, and $R_{\rm c,2}=5\times 10^{15}$ cm with 
 the linear density in the steady-state wind zones
 $w_1=2.1\times10^{17}$ g cm$^{-1}$ and 
 $w_2=2\times10^{16}$ g cm$^{-1}$. 
In addition, we plot the Thomson optical depth (Fig. 1) assuming  
 full ionisation of the CS gas 
 and the hydrogen abundance $X=0.4$.
The latter is estimated from the fact that the flux ratio of 
  He~II 4686 \AA\ to H$\alpha$ in the spectrum on 
 1998 March 6 is about unity (Fassia et al. 2001) and assuming 
 recombination case B.
 (The low hydrogen abundance may reflect mixing and mass loss).
Note, at $r=10^{15}$ cm the model yields            
 the Thomson optical depth $\tau_{\rm T}\approx 0.3$ 
 (Fig. 1), which is in a qualitative agreement with 
  $\tau_{\rm T}\approx 0.2$ derived by Lentz et al. (2001).

\begin{figure}
\centerline{\hspace{0.0cm}
\psfig{file=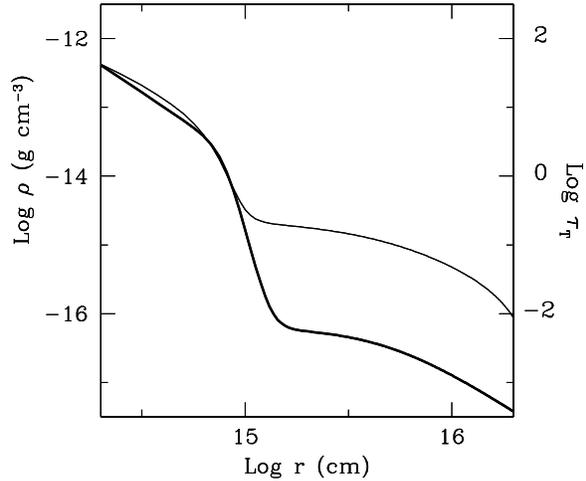,height=9 cm,angle=0}
}
\caption{
The density distribution of the CS gas in the model B (thick line).
Thin line shows the Thomson optical depth of CS gas. 
}
\label{fig1}
\end{figure}

\subsection{The opaque cool dense shell} 

The interaction of the freely expanding SN ejecta with the CS environment 
 produces a double shock structure with the outer (direct) shock, running in 
 the wind and the inner (or reverse) shock propagating inward the 
 SN ejecta (Nadyozhin 1981, 1985; Chevalier 1982).
If the CS wind is dense enough ($w\geq 3\times 10^{15}$ g cm$^{-1}$), 
 a cool dense shell (CDS) forms at the ejecta-wind interface
 due to the cooling and compression of the shocked ejecta in the 
 inner shock (Chevalier \& Fransson 1985, 1994).
In case of a very dense wind ($w\geq 10^{17}$ g cm$^{-1}$) 
 the outer shock during the initial epoch of 
 about one month may be radiative as well and contribute to the CDS formation.
The evidence for the CDS in SN~1998S is hinted by the
 'boxy' H$\alpha$ emission at the late epoch
   (Leonard et al. 2000; Gerardy 2000; 
 Fassia et al. 2001).
The blue slope of the H$\alpha$ boxy component, when
 approximated by the emission from a spherical shell,
 suggests the average CDS velocity  
 $5600 \pm500$ km s$^{-1}$ and 
 $5500 \pm500$ km s$^{-1}$ according to spectra
  on day 92 (Fassia et al. 2001) 
 and on day 140 (Leonard et al. 2000) respectively. 
 
Remarkably, the CDS 
  turns out opaque in the Paschen continuum for about 40--50 days.
This claim is supported by the following arguments.
The mass of the CS shell in the range of $10^{15}$ cm for the template wind  
 model is $\sim 0.1~M_{\odot}$.
After the CS shell is swept up by the SN ejecta with the density 
 in outer layers $\rho\propto v^{-\omega}$ one expects
 that the mass of the CDS should be $(\omega-4)/2$ times larger 
 (Chevalier 1982) or $M_{\rm s}\sim 0.2-0.8~M_{\odot}$ depending on 
 $\omega$, which lies in the range 8--20 (Chevalier \& Fransson
 1994).
We assume that the excitation temperature is 
 equal to the effective temeperature $T_{\rm eff}$ and approximate
 the bolometric luminosity evolution by the power law $L=L_0(t_0/t)^2$,
 where $L_0 = 10^{43}$ erg s$^{-1}$ and $t_0=40$ d, 
 which is compatible with the observed luminosity   
 during the period of 15--80 d after the discovery (Fassia et al. 2000).
With these remarks, the condition 
 $\tau_3(0.65\,\mu{\rm m})=1$ at H$\alpha$ or 
 $\tau_3=2$ at the Paschen edge 
 results in the following expression for 
 the time of the CDS transparency $t_1$ in the H$\alpha$ band 

\begin{eqnarray}
t_1=44\left(\frac{t_0}{40\,\mbox{d}}\right)^{0.5}
  \left(\frac{v}{6000\,\mbox{km/c}}\right)^{-0.5}
   \left(\frac{L_0}{10^{43}\,\mbox{erg/s}}\right)^{0.25}\nonumber\\
   \times(1+0.05\ln\,\phi)\quad \mbox {days},
\label{eq:t1}
\end{eqnarray}

\noindent where $v$ is the CDS average expansion velocity,
 $\phi\propto M_{\rm s}X(1-x)t_1^{-2}v^{-2}$ and 
 the numerical prefactor in Eq. (\ref{eq:t1}) is obtained for 
 $M_{\rm s}=0.5~M_{\odot}$, $X=0.4$, H ionisation degree
 $x=0$, $t_1=44$ d, and $v=6000$ km s$^{-1}$, which implies 
 $\phi=1$.
According to Eq. (\ref{eq:t1}) the 
 value of $t_1$ weakly depends on parameters, which gives us 
 a confidence that the CDS is actually opaque during initial 40--50 days.

Two important consequences follow from the fact that 
 SN ejecta is enshrouded in the opaque CDS.
First, the SN photosphere should reside in CDS at the epoch $t<t_1$, which
 matters to the observational test of the 
 deceleration dynamics (section 3).
Notably, the optically thick electron-scattering CS gas 
 does not affect the photospheric continuum, because CS shell 
 is effectively thin,
 {\em i.e.} the scattering is conservative.
Actually, for the typical density of the CS shell 
 $n_{\rm e}\sim 10^{10}$ cm$^{-3}$
 and temperature $T_{\rm e}\sim (2-3)\times10^4$~K the ratio 
 of the absorption to the extinction coefficient is $\epsilon\sim 10^{-3}$, 
 so the thermalization length  $1/\sqrt{\epsilon} \gg \tau_{\rm T}$.

The second point, but of a primary significance, is that 
 the opaque CDS prevents the emergence of strong 
  emission lines from the SN ejecta.
Moreover, strong emission lines are not expected from the opaque CDS either.
To clarify the latter point, note that the pressure equilibrium and the mass
 conservation result in a small width of the CDS (Chevalier \& Fransson 1994),
 $\Delta R/R\ll 1$, which leads to the small width of the 
 SN atmosphere residing in the CDS,
  $\Delta R_{\rm atm}/R<\Delta R/R\ll 1$. 
Since the early-time continuum is nearly 
 black-body, $I_{\nu}\approx B_{\nu}(T)$, and 
 the line source function in the CDS is limited by the inequality 
 $S_{\nu}\leq B_{\nu}(T)$, we conclude immediately 
 that the expected contrast of emission lines 
 formed in the opaque CDS is very small,  
  $\Delta F/F \approx (S_{\nu}/B_{\nu})(2\Delta R_{\rm atm}/R)\ll 1$.
It becomes thus clear, 
 why strong emission and absorption lines of SN~1998S ejecta 
 were not seen during  the period 10--40 days
 after the discovery, the observational 
 fact emphasised by Fassia et al. (2001). 
On the other hand, the opaque CDS does not preclude the emergence 
 of the strong emission lines from the extended CS shell.

To summarize, the qualitative 
 model for the early-time spectrum of SN~1998S 
 may be portrayed as a sharp black-body photosphere embedded 
 into the opaque
 electron-scattering CS shell, which is responsible 
 for the emission lines observed before March 12.

\begin{table*}
  \caption{Parameters of light curve models}
  \begin{tabular}{cccccccccccc}

Model  & $M$   &   $E$  & $R_0$ & $R_{\rm c,1}$ & $w_1$ & $w_2$ &
 $t_1$ & $t_{\rm c}$ & $R_{\rm s}$ & $T_{\rm eff}$ & $\tau_{\rm T}$\\

     & M$_{\odot}$ & $10^{51}$ erg  & \multicolumn{2}{c}{ $10^{14}$ cm}  &
\multicolumn{2}{c}{ $10^{16}$ g cm$^{-1}$} &
 \multicolumn{2}{c}{days} & $10^{14}$ cm & K & \\

 A &  5 & 1.1 & 2.4 & 8    & 21  & 2 & 47 & 19 & 5.0  & 21700 & 3.4 \\ 
 B &  5 & 1.1 & 2.4 & 10   & 18  & 2 & 47 & 22 & 5.0  & 21500 & 3.8 \\ 
 
\end{tabular}
\label{tab1} 
\end{table*}

\begin{figure}
\centerline{\hspace{0.0cm}
\psfig{file=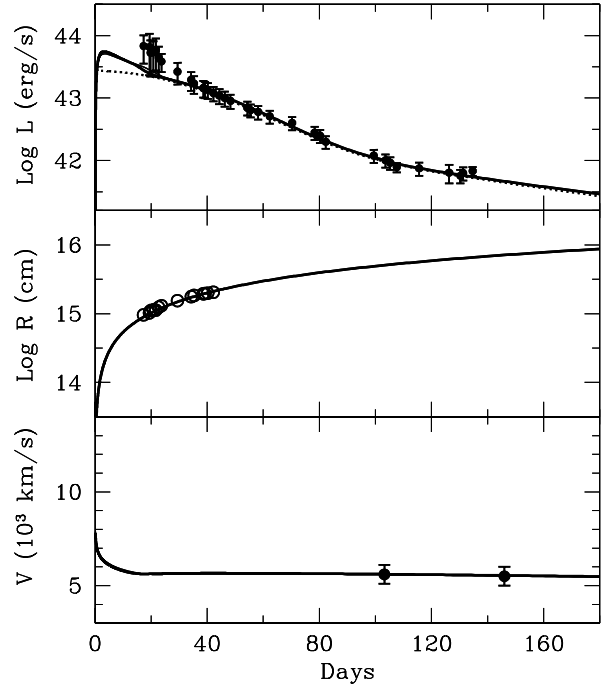,height=10cm,angle=0}
}
\caption{
Bolometric light curve of SN~1998S, 
 radius and velocity of the thin shell.
Upper panel: the light curve model A (see Table 1)
 shown by thin line, and B model (thick line) are 
 overploted on the empirical bolometric luminosity 
 (dots) from Fassia et al. (2000).
The light curve omitting CS interaction is shown by dotted line.
Middle panel: the radius of the thin shell for both
 models A and B; empirical photospheric radii (Fassia et al. 2000) 
 are shown by open circles.
Lower panel: the velocity of the thin shell 
 for A and B models with overploted estimates of the thin shell 
 velocity (see text). 
Note, both models produce similar curves in all the panels. 
}
\label{fig2}
\end{figure}

\section{The interaction and light curve model} 

Given the similarity of SN~1998S and SN~1979C photometric and 
 spectral properties (Liu et al. 2000), the  
 light curve of SN~1998S is likely of the same 
 nature as that of SN~1979C.
In line with the hydrodynamical model of the SN~1979C  
 (Blinnikov \& Bartunov 1993) we suggest that 
 the light curve of SN~1998S is the result of  
 an explosion of a red supergiant with an
 extended envelope,  $R\sim (1-10)\times 10^3~R_{\odot}$,
  and a moderate mass of ejecta $\sim 5~M_{\odot}$.
The radioactivity as well as the ejecta-wind interaction 
  are likely responsible for the late-time SN~1998S luminosity. 
One cannot rule out that 
 the ejecta-wind interaction may contribute to 
 the initial phase of the light curve as well.

The bolometric light curve will be simulated using a  
   linear composition of the analytical bolometric
  light curve of 'bare' ({\em i.e.} without wind) SN~II  
   (Arnett 1980, 1982) and the light curve powered by the ejecta-wind
    interaction (Chugai 1992).
The latter model computes the optical luminosity,
 produced as a result of reprocessing of the X-ray radiation from 
 the outer and inner shocks.
This model is upgraded here to include the Compton cooling of the
 postshock gas (Fransson 1982) and the reprocessing of 
  the X-ray radiation by the wind material as well.
The dynamics of the ejecta deceleration in a thin shell 
 approximation (Chevalier 1982) is solved numerically, 
 which yields the thin shell radius $R_{\rm s}(t)$ and its velocity 
 $v(t)$. 
The light curve computed in an approximation of 
 the instant radiation escape from the CS shell, is corrected 
 for the diffusion delay in the CS medium using Sobolev (1980) analytical 
 formula for the escaping luminosity.

The density profile of a freely expanding SN envelope 
 ($v=r/t$) used for the calculations 
 of the thin shell dynamics is a plateau with an outer 
 power law tale 

\begin{equation}
\rho=\rho_0(t/t_0)^{-3}[1+(v/v_0)^{\omega}]^{-1},
\end{equation}  
  
\noindent where $\rho_0$ and $v_0$ are defined by the SN mass ($M$) and  
 kinetic energy ($E$); $t_0$ is an arbitrary refference moment. 
We adopt a steep density gradient, $\omega=16$, to take into account 
 the effect of sweeping up the material of the extended envelope 
 into a dense shell
 during shock wave breakout phase (Grasberg et al. 1971; 
 Falk \& Arnett 1977).
The template CS density distribution (Fig. 1) is used.
The $^{56}$Ni with the mass of $0.15~M_{\odot}$ (Fassia et al. 2000)
 is assumed to be mixed within inner half of the SN mass.
The average SN ejecta opacity $k=0.2$ cm$^2$ g$^{-1}$ is adopted
 for the Arnett's model.

The list of the essential parameters of the light curve 
 model includes the ejecta mass $M$, kinetic energy 
 $E$, presupernova radius $R_0$, the wind parameters $w_1$ and $w_2$
 assuming the fixed shape of the density distribution of the SN ejecta and 
 the CS gas.
Exploring the parameter space led us to the optimal choice 
 of the SN parameters: $M=5~M_{\odot}$, 
 $E=1.1\times10^{51}$ erg and $R_0=2.4\times10^{14}$ cm. 
It should be stressed, that these values may be in some error 
 given a simplicity of our model.
The density in the inner CS shell ($w_1$) affects the very early 
 phase $<30$ d of the light curve, 
 although, we are not able to derive $w_1$ reliably from 
 the light curve analysis, since the Arnett's analytical model 
 underestimates the early luminosity (Arnett 1980). 
The light curve is insensitive to 
 the outer wind density unless 
 $w_2$ substantially exceeds 
 $2\times 10^{16}$ g cm$^{-1}$.

In Fig. 2 two models of the light curve are shown for
 input parameters given in Table 1 and assuming 
 that the explosion occured on 1998 February 24.7 UT, the day 
 after the last pre-discovery observation.
This choice is consistent with 
 the evolution of the photospheric radius (Fig. 2).
The template CS density distribution (Fig. 1) is used for 
 the model A, while the model B differs by the slightly higher cutoff radius
  $R_{\rm c,1}$.
The parameter $w_1$ is found from the 
 modeling of the light curve, photospheric radius, CDS velocity and 
  the line broadening effect (section 3) using an itterative procedure.
Derived values shown in Table 1 include 
 the moment $t_1$, when the optical depth of the CDS in the Paschen continuum 
  at H$\alpha$ wavelength $\tau(0.65\mu{\rm m})=1$, the time $t_{\rm c}$, 
  when the outer shock becomes adiabatic,  and three important
  values on March 6: the radius of the thin shell
 ($R_{\rm s}$), which coincides with the CDS, 
  effective temperature ($T_{\rm eff}$),
 and Thomson optical depth ($\tau_{\rm T}$) of the wind.
The optical depth of the CDS in the Paschen continuum is calculated   
 assuming LTE for $T=T_{\rm eff}$ 
 and the number density of the CDS defined by
 the isobaric condition, {\em i.e.} with   
 the thermal pressure equal to the dynamical pressure $\rho v^2$,
 where $\rho$ is the preshock wind density and $v$ is the thin 
 shell velocity.
Models A and B produce similar results being consistent with 
 the empirical light curve, photospheric radii and CDS velocity. 
Some luminosity deficit before day 30 might be related to
 the mentioned fact that in the analytical light curve the initial 
 luminosity peak is undeproduced (Arnett 1980).
Yet, given large uncertainties of data, the difference 
 between the model and the empirical light curve is not dramatic.  
Note, the factor of two increase in the luminosity corresponds to 20\%
 higher value of $T_{\rm eff}$.

The CDS is opaque during initial $\approx 47$ days in 
 agreement with the previous estimates (section 2). 
Taking into account  the six days lag between the explosion 
 and the discovery we conclude 
 that the opaque CDS gets transparent at $\approx 40$ d after 
 the discovery. 
This is consistent with the observational fact that 
 strong broad emission and absorption lines from SN ejecta 
 appear only about day 40 after the discovery (Fassia et al. 2001).
Remarkably, the outer shock is radiatiave in 
 both models for $\approx 20$ days.
As a result 
 on March 6 with $\approx 9$ days passed after the explosion 
 the outer postshock gas should cool very quickly resulting in a 
 very thin outer postshock region, {\em i.e.} $\Delta R/R < 0.1$.

\section{Line profile calculations}

We consider only the H$\alpha$ line, although 
 a similar approach may be applied to any other line, including 
 WR lines, observed in early-time spectra of SN~1998S.
The model suggests a sharp photosphere with a radius $R_{\rm p}=R_{\rm s}$ 
 in a fully ionised 
 isothermal CS gas ($r>R_{\rm p}$) 
  with an electron temperature $T_{\rm e}=T_{\rm eff}$ (Table 1).
Any effects related to the outer shock are omitted; they are of a minor 
 importance for the line profile.
The wind velocity law     
 $v=v_{\rm max}(R_{\rm p}/r)^2 + v_{\rm min}$ is taken to mimic  
  a radiative acceleration of the wind by the SN radiation.
We adopt $v_{\rm min}=40$ km s$^{-1}$ 
 for the distant wind (Fassia et al. 2001) and 
 $v_{\rm max}=1000$ km s$^{-1}$ at the photosphere, a rough guess 
 based upon the extrapolated behavior of the maximal 
 velocities of the fast
  CS gas on March 20 and April 8 (Fassia et al. 2001).
The result, actually, is not very much sensitive to $v_{\rm max}$
 unless it significantly exceeds 1000 km s$^{-1}$.
The wind emissivity in  H$\alpha$ is assumed to scale 
 as a recombination 
  emissivity $j=C_{\rm em}(h\nu/4\pi)\alpha_{32}n_{\rm e}n(\mbox{H}^{+})$
  (erg s$^{-1}$ cm$^{-3}$ ster$^{-1}$), 
   where  $\alpha_{32}$ is the effective recombination coefficient for 
 the H$\alpha$ emission (case B), while  $C_{\rm em}$ is an emission
  correction factor, which allows for uncertainties 
  in the distance, exctinction, electron temperature, hydrogen abundance,
  as well as for a wind clumpiness and   
 a possible deviation from the recombination case B.
The scattering in the wind is essentially conservative, but
 the photons struck the photosphere will be considered lost.
The resonance scattering is neglected. 
This approximation is justified by our simulations, which do not show 
 a dependence on the resonance scattering at the considered epoch.

The emergent line spectrum is computed using the Monte Carlo technique.
The photon frequency at each scattering is randomly chosen adopting 
 the symmetric angle-averaged frequency 
 redistribution function (cf. Mihalas 1978).
Note, computations with both
  angle-averaged and angle-dependent frequency redisribution 
 functions are mutually consistent with a high precision even in the 
 case of $\tau_{\em T}\approx 1$ as Hillier (1991) has shown 
 in his study of electron-scattering effects in WR stars.
The relativistic correction to 
 the Doppler effect and Compton recoil produce a net 
 blue shift  
 $\Delta \nu/\nu = N_{\rm s}(3kT_{\rm e}-h\nu)/mc^2$ (Weymann 1970), 
 where $N_{\rm s}$ is the average number of scatterings.
This effect, although small, is included in the profile modeling.

\begin{table}
  \caption{Parameters of demonstration models for line profile}
  \begin{tabular}{ccccc}

Model & $T_{\rm e}$ & $w_1$ & $\tau_{\rm T}$ \\

    & K  & $10^{16}$ g cm$^{-1}$ & \\

 D1 & 25000  &  13   &    4.0  \\
 D2 & 25000  &  1.3  &    0.4  \\
 D3 & 25000  &  16.2  &    5.0  \\
 D4 & 30000  &  13   &    4.0  \\

\end{tabular}
\label{tab2} 
\end{table}

The H$\alpha$ line broadening effect in the case of the high Thomson 
 optical depth is illustrated 
 in Fig. 3 with model parameters given in Table 2, 
 which presents the electron temperature, the linear wind density 
 $w_1$, and the Thomson optical depth.
In all the cases the photospheric radius is 
 $R_{\rm p}=4\times10^{14}$ cm, 
 the wind velocity is $u=40+1000(R_{\rm p}/r)^2$ km s$^{-1}$
  and the CS density distribution is
 defined by Eq. (\ref{eq:cs}) with $w_2=0$, 
 $R_{\rm c,1}=9\times10^{14}$ cm, $s_1=-2$, $p_1=10$. 
All the profiles are normalized to unity at maximum. 
A gaussian smoothing with FWHM=6.7 \AA\
 is applied henceforth to the computed profile to allow for the 
 finite resolution 
 in the spectrum on 1998 March 6 (Fassia et al. 2001).

Results show that the template model D1 with the Thomson optical depth 
 $\tau_{\rm T}=4$ produces strong nearly symmetric 
 electron-scattering wings. 
It should be emphasised that the line profile from the expanding 
 electron scattering wind with the opaque core 
 is the result of the combination of 
 several factors including, (i) single scattering broadening 
 related to the outmost layers with $\tau_{\rm T}\leq 1$; 
 (ii) broadening due to multiple scattering, which may be interpreted  
  as a diffusion in the frequency space; (iii) the 
 redshift acquired in the course of multiple scattering in the 
 expanding medium; 
 (iv) the photon absorbtion by the opaque photosphere.
It should be emphasised that the resulting redshift essentially 
 depends on the wind expansion kinematics. 
The model D2 with ten times lower optical depth compared to the model D1 
 clearly shows the blue shift due to the occultation of the wind 
 by the photosphere and the enhanced red wing due to the 
 expansion effect. 
A primary reason, why the expansion effect is not so apparent in the 
 model D1 relates to the adopted wind kinematics with the velocity 
 dropping outward. 
In the optically thick situation the contribution of the outer 
 slow expanding gas dominates, which thus account for the small 
 expansion effect in the model D1.
Obviously, this effect does not operate in the transparent case D2, 
 where fast inner scattering wind essentially contributes in the 
 line profile.  
Note, the mentioned bias to the red wing in the model D2 very much 
 resembles the appearence of electron-scattering wings   
 in WR stars (Auer \& van Blerkom 1972; Hillier 1991).
The model D3 with $\tau_{\rm T}=5$ has markedly stronger  
 wings compared to the model D1 in a wide range of velocities.
The  20\% higher electron temperature (model D4) 
 slightly increases high velocity wings.
Such an effect is practically on the verge of detectability. 
In this regard we note, that the uncertainty of $T_{\rm eff}$ 
 in the light curve model is of $\approx 20\%$, which is thus 
 insignificant for the line modeling.

\begin{figure}
\centerline{\hspace{0.0cm}
\psfig{file=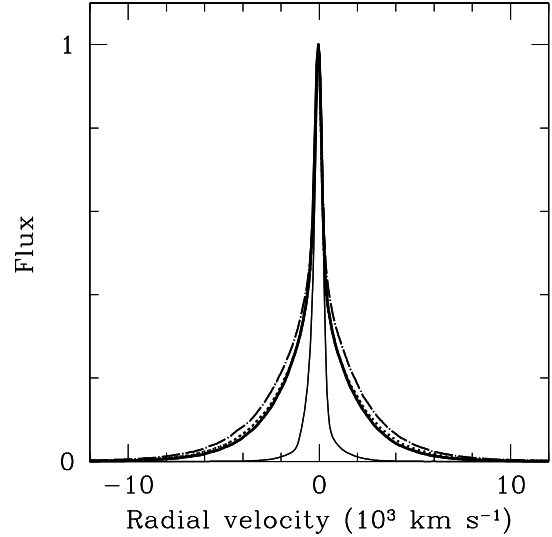,height=10cm,angle=0}
}
\caption{
The line broadening effect for different parameter sets (see Table 2).
Shown are the template model D1 (thick solid line),
 model D2 with small Thomson optical depth 
 (thin line), model D3 with higher optical depth (dot-dashed line), 
 and model D4 with higher electron temperature (dotted line).
}
\label{fig3}
\end{figure}

The spectrum of H$\alpha$ on 1998 March 6 
 (Fassia et al. 2001) with two 'best' fit versions of the 
 line profile based upon models A and B are ahown in Fig. 4.
To make the plot the emission lines are placed upon the 
 appropriate continuum.
As noted before (section 3),   
 parameters of models A and B are found 
 using itterative procedure to fit simultaneously 
 the light curve, photospheric radius, expansion 
 velocity and the line profile.  
Both models reproduce the observed profile with the
 emission correction factor $C_{\rm em}$ being close to unity
 (1.36 and 1.13 for model A and B respectively).
Surprisingly enough that the model A, which only slightly differs 
 from the model B, shows a noticeable blue shift.
This is related to the smaller ratio $R_{\rm c,1}/R_{\rm p}$, {\em i.e.} 
 higher compactness of the CS shell. 
Note, the higher blue shift would be inconsistent with the 
 observed profile, which implies that $R_{\rm c,1}\geq 8\times 10^{14}$ cm.
On the other hand with the Thomson optical depth of CS gas at the radii
 $r>10^{15}$ cm as low as $\tau_{\rm T}\sim 0.2$ (Lentz et al. 2001) 
 the outer boundary of the CS shell cannot significantly exceed  
  $10^{15}$ cm.
Therefore we conclude that $R_{\rm c,1}\approx 10^{15}$ cm is an 
 optimal estimate for the cutoff radius of the inner CS shell.
 
With the single studied epoch we cannot rule out 
 that the density distribution at $r<R_{\rm c,1}$ may 
 deviate from the law $s_1=-2$.
For instance, the profile may be fitted  
 assuming $s_1=-2.5$ and $R_{\rm c,1}=10^{15}$ cm 
  with similar value of the optical depth,
   $\tau_{\rm T}\approx 3.5$.
Other uncertainties of the model include  
 the clumpiness, and  asymmetry of CS gas.
The clumpiness would result in the decrease of the Thomson 
 optical depth compared to the smooth distribution. 
The global asymmetry of the CS environment (if any) would 
 suggest that our results must be refered to some equivalent 
 spherical model.
At any case, uncertainties are not able to change 
 the major result, that the broad line wings on March 6
 require the Thomson optical depth of the CS shell in the range 
 of 3--4.

\begin{figure}
\centerline{\hspace{0.0cm}
\psfig{file=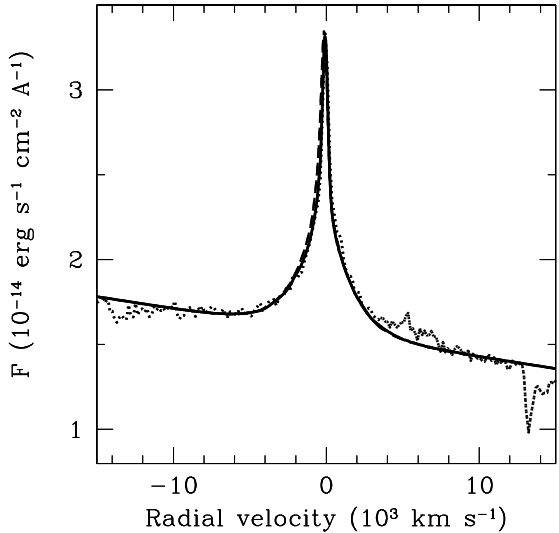,height=10cm,angle=0}
}
\caption{
The H$\alpha$ profile in SN~1998S spectrum on 1998 March 6.
The model A (dashed line) and model B (solid line) are overploted on 
 obsrevations by Fassia et al. (2001).
}
\label{fig4}
\end{figure}

\section{Discussion}

I have argued that the broad narrow-topped  
 emission lines in the early-time spectra of SN~1998S are  
 the result of the emission and Thomson scattering of
 line photons in the opaque CS shell with the outer radius 
 of $\approx 10^{15}$ cm.
It was demonstrated that the CS density distribution appropriate 
 to the line broadening 
 is consistent also with the constraints imposed by the light curve,
 photospheric radius and velocity of the CDS.
It was shown that the CDS is opaque in the Paschen continuum 
 during about 40--50 days after the SN explosion.
The absence of strong emission and absorption lines from SN ejecta 
 during 40 days after the SN discovery (Fassia et al. 2001) is 
 explained as a direct consequence of the opaque CDS.
Both the electron-scattering CS environment and 
 the opaque CDS, thus, may produce pronounced observational effects 
 in SN~IIn. 

Despite an obvious success in explaining the spectral evolution of 
 SN~1998S, the concept of the opaque CDS faces a problem in 
  accounting for the weak broad absorptions of H, HeI and Si II, 
  which are noticeable in SN~1998S spectra before day 40 (Fassia et al. 2001).
A plausible explanation for these lines might be 
   that the CDS is not a perfect sphere,
   but is instead 'punched' {\em i.e.} has 'holes'.
In such a case the absorbing gas may be identified with 
 the unshocked gas of SN ejecta seen through the holes in the opaque CDS.
A punched structure of the CDS might be caused by the Rayleigh-Taylor  
 instability or by a global wind asymmetry.
The latter is implied by the detection of the intrinsic 
 polarization of SN~1998S on 1998 March 7 (Leonard et al. 2000).
For instance, 
 the asymmetric opaque CDS might have an appearence of a wide equatorial 
 belt with two transparent poles, one of them 
 lying in the near hemisphere.
 
In regard to the umbiguous interpretation of the spectropolarimetric 
 data (Leonard et al. 2000) the following property of our model 
 may be relevant. 
Specifically, the model predicts that the narrow component 
 in the early-time spectrum must consist mostly of the non-scattered 
 radiation, while the broad wings are the result of multiple 
 scattering.
With this property more consistent is the interpretation II 
 (polarized broad lines and unpolarized narrow lines) suggested by 
 Leonard et al. (2000). 
 
The origin of the inner CS shell with the outer radius of 
 $\approx 10^{15}$ cm is intriguing.
If it was created by the slow red supergiant wind with the 
 velocity $u=10$ km s$^{-1}$, then the age of this CS shell was
 30 yr at the explosion, which cannot be readily associated 
 with any specific time scale of the presupernova.
On the other hand, the velocity of the matter ejection might be 
 as high as 100 km s$^{-1}$, 
 which is still tollerated by the spectral resolution of the 
  narrow component (Fassia et al. 2001).
In that case the age of the CS shell 
   was only 3 yr, which reminds the time scale 1--10 yrs for the Ne and O 
   burning in cores of $12-20~M_{\odot}$ stars (Heger 1998).
Woosley (1986) noted that the Ne burning in degenerate cores of 
  $\sim 11~M_{\odot}$ stars may result in the violent mass loss
  roughly 1 yr prior to core collapse.
Remarkably, the dense shell of a similar size ($\sim 10^{15}$ cm) 
 has been invoked to 
 account for the early-time line behavior in SN~1983K (Niemela et al. 1985),
 which is another SN~II with WR features, and in type IIn SN~1994W 
 (Sollerman et al. 1998).

It would be interesting to search for signatures of  
 broadening of CS lines due to scattering by thermal electrons in 
 other SN~IIn spectra. 
To distinguish between the case of narrow 
 lines on top of the broad lines emitted by SN ejecta, likewise in {\em e.g.} 
 SN~1988Z (Filippenko 1991; Stathakis \& Sadler 1991) and the case of 
 narrow lines with CS electron-scattering wings,
 the following properties of the early-time spectra of 
  SN~1998S must be stressed: 
 (i) the broad wings are smooth;
   (ii) strong emission lines have similar profiles;
 (iii)  the continuum is smooth and close to black-body.
We deliberatelly do not emphasize the wing symmetry, since in the 
 case of a low Thomson optical depth and a high wind velocity
 the red wing may become stronger due to the CS expansion effect and   
 the resonance scattering.
In this regard of particular interest are 
 SN~1994Y, SN~1994W, and SN~1994ak, reviewed by Filippenko (1997).
Their early-time spectra show narrow lines with smooth broad wings 
 on smooth continua. 
These properties are very suggestive of the CS electron-scattering 
 origin of line wings and the presence of the opaque CDS in these 
 supernovae.

\section*{Acknowledgments}

I am grateful to 
  Peter Lundqvist, Peter Meikle and Alexandra Fassia for 
 discussions of different issues of SN~1998S 
 and for a creative aura during my visit to Saltsj\"{o}baden and 
 Imperial Colledge.
 
The study was supported by RFBR grant 01-02-16295.

\end{document}